\documentstyle[epsf,11pt]{article}
\newcommand{\bea}{\begin{eqnarray}}
\newcommand{\eea}{\end{eqnarray}}

\begin{document}
\begin{titlepage}

\thispagestyle{empty}

\vspace{0.2cm}

\title{{\normalsize \hfill HIP-2001-19/TH}\\[20pt]
Radions in a $\gamma\gamma$ collider 
\footnote{The authors thank the Academy of Finland
(project number 163394 and 48787) for financial support.}}
\author{M. Chaichian$^{a,b}$, K. Huitu$^a$, A. Kobakhidze$^{a,b}$ and
Z.-H. Yu$^{a,b}$\\
$^a$Helsinki Institute of Physics\\
$^b$Department of Physics, University of Helsinki\\
P.O.Box 9, FIN-00014 Helsinki, Finland}
\date{}
\maketitle

\vspace*{2truecm}

\begin{center}\begin{minipage}{5in}

\begin{center} ABSTRACT\end{center}
\baselineskip 0.2in

{We study the resonance production of radions via $\gamma\gamma$ fusion
in the Randall-Sundrum model.
We find that the cross section of the process 
in the $\gamma\gamma$ mode of a linear collider (LC) can be of similar size 
as in the $e^+e^-$ collision mode, if the radion is heavy. 
We consider the possible curvature-Higgs mixing in the model,
and we find that the mixing should be constrained in order to avoid 
an unphysical state.  
The process $\gamma\gamma \rightarrow \phi$ is the main source of
radions  at 
LC near the conformal limit. We consider both the decay modes of
radion with curvature-Higgs mixing and without.
Our results show that the radion with mass below 800 GeV could be 
detected at LC with any mixing parameter and vacuum expectation value
of the radion around 1 TeV.} \\

\vskip 5mm

\end{minipage}
\end{center}
\end{titlepage}

\eject
\rm
\baselineskip=0.25in

\begin{flushleft} {\bf 1. Introduction} \end{flushleft}

\noindent
One of the most fascinating outcome of the recently proposed brane
World 
scenarios 
\cite{s1, s2} is a possibility to probe quantum gravity and entire 
structure of space-time 
at high energy colliders. The main theoretical motivation behind 
these models is a  
solution to the familiar problem of the observed hierarchy of scales, 
such as $M_{EW}/M_P<<1$, 
where $M_{EW}$ and $M_P$ are the electroweak and Planck scales, respectively. 

An interesting possibility  has been 
proposed by  Randall and Sundrum 
(RS) within the model with 5-dimensional AdS space-time compactified on a 
$S^1/Z_2$ orbifold with 
two 3-branes stucked at the orbifold fixed-points (see the first paper 
in \cite{s2}). The background geometry 
has been found to be non-factorizable with a warp factor 
in the metric being an exponentially decreasing function 
of the fifth compact coordinate. Such a geometry provides the
localization 
of zero-mode graviton 
on a Planck brane  and exponential hierarchy between mass scales 
as seen by observers on different branes. 
Thus, contrary to the proposal of ref. \cite{s1}, even small distance 
between two branes, determined by the radius of extra dimension, can 
explain the large 
hierarchy between $M_{EW}$ and $M_P$ 
if the visible world is localized on a TeV brane.

The size of compact extra dimension is given by a vacuum expectation 
value (VeV) of a certain scalar field called 
modulus (radion). Usually the moduli have runaway VEVs. Thus, the radion 
stabilization (see {\it e.g.} \cite{s3,s3a}) in \cite{s1, s2} is a 
rather crucial point not only for the 
desired explanation of the hierarchy problem but also for obtaining 
consistent 4-dimensional gravity and 
cosmology on the visible TeV brane. 
Remarkably, the stabilized radion in the RS model is typically 
lighter than 
the low-lying Kaluza-Klein modes 
of graviton \cite{s3, s4}. Thus the radion might be the first 
state accessible in the 
high-energy experiments and study 
of its collider phenomenology will be important in order to test the model.

Recently many works have discussed the phenomenology
of radions \cite{s4,s5,s5a,s5b,new1}. 
We'll study in this paper a specific model with curvature-Higgs 
mixing \cite{s6,new2}. At the two-derivative
level scalars and gravity are coupled in the visible
brane, 
\bea
S=-\xi \int d^4 x \sqrt{-g_{vis}} R(g_{vis}) H^+ H,
\eea
where the Ricci scalar $R(g_{vis})$ corresponds to the induced 
four dimensional metric on the visible brane and $H$ is the electroweak 
Higgs boson. This term will introduce mixing between radion and
Higgs in RS model, and will modify the phenomenology of radion
strongly when compared to the unmixed case \cite{s6}. 
Especially near the conformal limit, $\xi = 1/6$,  radion coupling to the
Standard Model fields $W$, $Z$ and fermions will be strongly suppressed,
while gluon and photon interactions with radion will be only partially 
supressed \cite{s6}. As a result, detecting radion at a Linear Collider (LC)
directly in $e^+e^-$ collision will become difficult and 
production of radion by $\gamma\gamma$ fusion will be significant. 

With the advent of new collider techniques, highly coherent laser
beams can be produced by back-scattering with high luminosity
and efficiency at the $e^{+}e^{-}$ colliders \cite{s7}.
A $\gamma\gamma$ resonance can be probed over a wide
mass region.
Thus the $\gamma\gamma$ collisions can produce heavier radions than
the corresponding $e^{+}e^-$ collisions.
Recently  in \cite{new1} the radion production in $\gamma\gamma$ colliders
without mixing of radion and Higgs was discussed.

In section 2, we will discuss the coupling of radion to SM fields
in the case of the curvature-Higgs mixing.
In section 3 the numerical results are presented.
Possible experimental signals are considered in section 4, where the
decay widths are presented.   
The conclusions are given in section 5 and some details of
the expressions are listed in the appendix.

\begin{flushleft} {\bf 2. Couplings of radion to the SM fields} 
\end{flushleft}

\noindent
In RS model \cite{s2}, the single extra dimension is compactified
on a $S^1/Z_2$ orbifold with coordinate $y$, $y {\cal 2} [-\pi,\pi]$.
The metric in the model is given by\footnote{Strictly speaking, 
inclusion of the Higgs-curvatute mixing term (1) may generally 
modify the background solution (2.1). Say, after the electroweak Higgs boson  
acquires a VEV, $H=\frac{1}{\sqrt{2}}(h(x)+v)$, the 4-dimensional
induced 
curvature will also appear 
in the effective Lagrangian affecting the background solution (2.1) 
\cite{n1}. Here we will 
ignore such a complication.} 
$$
ds^2=e^{-2 k r_c |y|} \eta _{\mu\nu} dx^{\mu}dx^{\nu} - r_c^2
dy^2. 
\eqno {(2.1)}
$$
where $1/k$ is the AdS curvature radius. 
\par
Following Refs. \cite{s5,s6}, we define a radion field
as 
$$
\phi=\Lambda_{\phi} e^{-k\pi(T-r_c)}
\eqno {(2.2)}
$$
where $M_5$ is the Planck scale of the fundamental 5-dimensional
theory, and $\Lambda_{\phi}$ is the VEV of the radion. 
$\Lambda_{\phi}=
\sqrt{24M_5^3/k} \,e^{-k\pi r_c}$, which can be of the order of TeV.
Integrating out the fifth dimension, one obtains the four 
dimensional effective action as \cite{s5,s6}
$$
S=\int d^4x \sqrt{-g} \left[\frac{2M_5^3}{k} \left(1-
\frac{\phi^2}{\Lambda_{\phi}^2} e^{-2k\pi r_c}\right) R +
\frac{1}{2} \partial _{\mu} \phi \,\partial ^{\mu} \phi-
V(\phi)+\left(1-\frac{\phi}{\Lambda_{\phi}}\right)T_{\mu}^{\mu}\right],
\eqno{(2.3)}
$$
where $V(\phi)$ is the potential stabilizing the radion field.
\par
We will concentrate on $T^{\mu}_{\mu}$ terms which will couple
to radion, as given in \cite{s6,new2},

$$
T^{(1)\mu}_{\mu}=6\xi v \Box h - 3\xi v^{2}/\Lambda_{\phi} \Box \phi,
\eqno{(2.4)}
$$

$$
T^{(2)\mu}_{\mu}=(6\xi-1)\partial _{\mu}h \,\partial^{\mu}h+
6 \xi h \Box h+ 2m_h^2 h^2+m_{ij} \bar{\psi}_i \psi_j-M_V^2 V_{A\mu}
V_A^{\mu},
\eqno{(2.5)}
$$
where $v$ is the VEV of Higgs.
We will neglect terms of stress-energy tensor containing three or 
more fields in this paper. 

$T^{(1)\mu}_{\mu}$ term induces a kinematic
mixing between Higgs and radion.  After shifting $\phi\rightarrow \phi +
\Lambda_{\phi}$, the Lagrangian containing bilinear terms of radion
and Higgs is obtained as
$$
{\cal L} = - \frac{1}{2}  \phi [(1-6\xi \gamma^2) \Box + m_{\phi} ^2 ] 
\phi
- \frac{1}{2} h (\Box + m_{h}^2 ) h - \frac{6 \xi v}{\Lambda_{\phi}}
\phi \Box h.
\eqno {(2.6)}
$$
Here $m_{\phi}$ is a mass parameter in $V(\phi)$ and 
$\gamma=v/\Lambda_{\phi}$. 
After diagonalization, the fields should be redefined as
$$
\phi= a \phi^{'} + b h^{'},
\eqno {(2.7)}
$$
$$
h= c \phi^{'} + d h^{'},
\eqno {(2.8)}  
$$
where 
$ a=\cos \theta /Z$; $b=-\sin \theta/Z$;
$c=\sin \theta-6\xi \gamma/Z \cos \theta$ and
$d=\cos \theta +6 \xi \gamma /Z \sin \theta$, 
with
$Z^2=1-6\xi \gamma^2 (1+6\xi)$ and
the mixing angle $\theta$ is given by
$$ \tan 2 \theta = 12 \xi \gamma Z \frac{m_h^2}{m_h^2(Z^2-36\xi^2\gamma^2)-
m_{\phi}^2}.
\eqno {(2.9)}  
$$ 
Our results agree with
those in Ref. \cite{s6} (with $\xi \gamma <<1$) and in Ref. \cite{new2}. 
{}From Eq. (2.7-8), we see clearly the
constraints $-(1+\sqrt{1+4/\gamma^2})/12 \le \xi \le 
(\sqrt{1+4/\gamma^2}-1)/12$, just as in Ref. \cite{new2}.

The new fields $\phi^{'}$ and $h^{'}$ are mass eigenstates with masses
$$
m_{\phi^{'}}^2= c^2 m_h^2 + a^2 m_{\phi} ^2
\eqno {(2.10)}
$$
$$
m_{h^{'}}^2= d^2 m_h^2 + b^2 m_{\phi} ^2 
\eqno {(2.11)}
$$
\par
The interaction Lagrangian of $\phi$ and $h$ with fermions and 
massive gauge bosons,
$$
\begin{array}{lll}
{\cal L} =-\frac{1}{v} (m_{ij} \bar{\psi}_i \psi_j - M_V^2 V_{A\mu}
V_{A}^{\mu}) \left[h+ \frac{v}{\Lambda_{\phi}} \phi\right]
\end{array}
\eqno{(2.12)}
$$
can be transformed to the coupling of mass eigenstates $\phi^{'}$ and
$h^{'}$ to fermions and massive gauge bosons as
$$
\begin{array}{lll}
{\cal L}=-\frac{1}{\Lambda_{\phi}} (m_{ij} \bar{\psi}_i \psi_j - M_V^2
V_{A\mu} 
V_{A}^{\mu}) [a_{34} \frac{\Lambda_{\phi}}{v} h^{'} + a_{12} \phi^{'}],
\end{array}
\eqno{(2.13)} 
$$
where $a_{12}= a+c/\gamma$ and 
$a_{34}=d+b\gamma$.
The coefficients $a_{12}$ and $a_{34}$ give directly the strength of
the corresponding interaction when compared to the case with no
mixing.

As pointed out in Ref. \cite{s6}, $a_{12}$ can be approximately
zero in the conformal limit $m_h=0, \xi=1/6$ when
$\Lambda_{\phi}>>v$. 
Thus the associated production in $e^+e^-\rightarrow Z\phi' $
becomes inaccessible at conformal limit and other production 
mechanisms should be considered.

The coupling of the radion to two Higgs bosons depends on
$V(\phi)$ and mixing of radion and Higgs. 
Neglecting the radion
self-coupling in $V(\phi)$, we can get the vertex of $h^{'}h^{'}\phi^{'}$
as
$$
\begin{array}{lll}
&&\frac{1}{\Lambda_{\phi}} (2 m_h^2 a d^2 h^{'2} \phi^{'} - a d^2 \phi^{'} 
\partial_{\mu} 
h^{'}
\partial ^{\mu} h^{'} (1-6\xi)  
+ 6\xi ad^2 (h^{'} \Box h^{'}) \phi^{'}\\
&&+ 4 m_h^2 b c d \phi^{'} h^{'2}-
2bcd h^{'} \partial_{\mu} \phi^{'} \partial^{\mu} h^{'} (1-6\xi)+
6bcd \xi h^{'} (\phi^{'}\Box h^{'}+h^{'} \Box \phi^{'})).
\end{array}
\eqno{(2.14)}
$$

In addition to the tree-level $T^{\mu}_{\mu}$, there is also
a trace anomaly term for gauge fields \cite{s8}.  The effective vertex
is given by
$$
[(\frac{1}{\Lambda_{\phi}} (a b_3 - \frac{1}{2} a_{12} F_{1/2}(\tau_t))
\phi^{'}+\frac{1}{v} (\frac{v}{\Lambda_{\phi}} b b_3 - 1/2 a_{34} F_{1/2}
(\tau_t)) h^{'}] \frac{\alpha_{s}}{8\pi} G_{\mu\nu}^a G^{\mu\nu a}
\eqno{(2.15)}
$$
for radion to gluons and 
$$
\begin{array}{lll}
&&[(\frac{1}{\Lambda_{\phi}} (a (b_2+b_Y) - a_{12}
(F_1(\tau_W) +\frac{4}{3} F_{1/2}(\tau_t))
\phi^{'}+\\
&&\frac{1}{v} (\frac{v}{\Lambda_{\phi}} b 
(b_2+b_Y) - a_{34} (F_1 (\tau_W) + \frac{4}{3} F_{1/2}
(\tau_t)) h^{'}] \frac{\alpha_{EM}}{8\pi} F_{\mu\nu} F^{\mu\nu}
\end{array}
\eqno{(2.16)}
$$
for radion to two photons, where $b_3=7$ is the QCD $\beta$-function
coefficient and $b_2=19/6, b_Y=-41/6$ are $SU(2)\times U(1)_Y$
$\beta$-function coefficients in the SM. $F_1$ and $F_{1/2}$ are form
factor from loop effects, which will be given in detail in appendix.
We can see from Eq.(2.14,15) that the vertex can remain nonvanishing
in the conformal limit. 
There will be significant terms when we consider physics near the
conformal limit.

\begin{flushleft} {\bf 3. Radion production} \end{flushleft}

\noindent
The radion production cross sections via $\gamma\gamma$ fusion will
be given in this section. 
When the mixing becomes strong, it will become less clear which is
Higgs and which is radion. Since the mixing matrix of radion and Higgs 
is not unitary, we will always consider 
$\phi^{'}$ as radion and $h^{'}$ as Higgs in the following calculations.  

\begin{figure}[t]
\leavevmode
\begin{center}
\mbox{\epsfxsize=8.truecm\epsfysize=6.truecm\epsffile{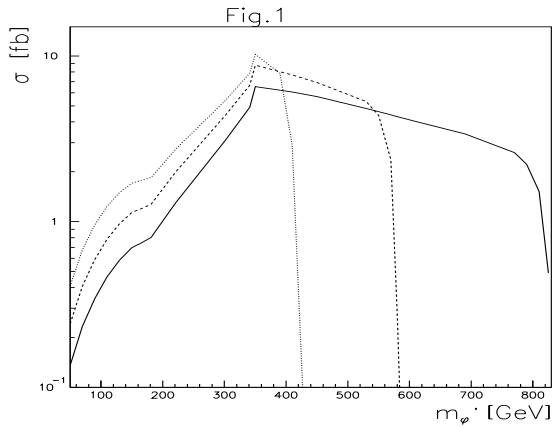}}
\end{center}
\caption{\label{fig1a} The cross section of $\gamma\gamma \rightarrow
\phi^{'}$ as a function
of $m_{\phi^{'}}$ with $\Lambda_{\phi}=1$ TeV, $m_h$=150 GeV and
$\xi=0$.
The solid line corresponds to the center of mass energy 1 Tev,
the dashed line to 700 GeV and the dotted line to 500 GeV.}
\end{figure}
In Fig.1, we show the cross section of $\gamma\gamma \rightarrow
\phi^{'}$ as a function of the mass of radion $m_{\phi^{'}}$ at 
c.m. energy $\sqrt{s}$=500 GeV, 700 GeV, and 1 TeV, respectively. 
In Fig. 1 radion is not mixed with Higgs, {\it i.e.} 
we set $\xi=0$, $m_h$=150 GeV, and $\Lambda_{\phi}=1$ TeV. 
Compared with the results of Ref. \cite{s5b}, where the
cross section of radion production from $e^+e^-$ mode at a
Linear Collider (LC) is considered, we find that the cross sections 
for heavy radions (with mass above 500 GeV) 
are of similar magnitude.
The process should be considered as an important way to produce radions
at LC, when there is no mixing. 
Our calculations agree with  the results of Ref. \cite{new1}
for this case.

\begin{figure}[t]
\leavevmode
\begin{center}
\mbox{\epsfxsize=6.truecm\epsfysize=6.truecm\epsffile{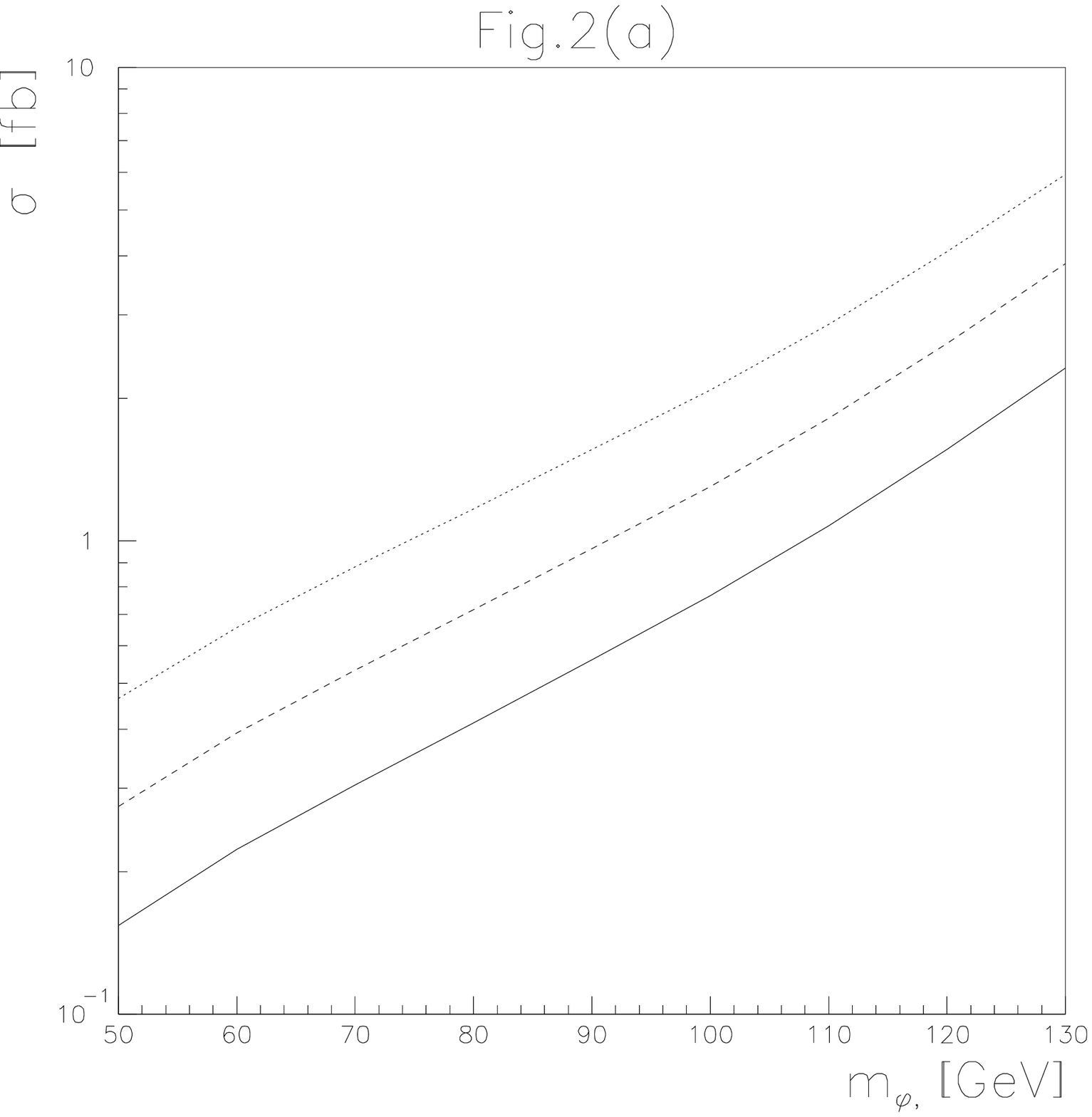}}
\mbox{\epsfxsize=6.truecm\epsfysize=6.truecm\epsffile{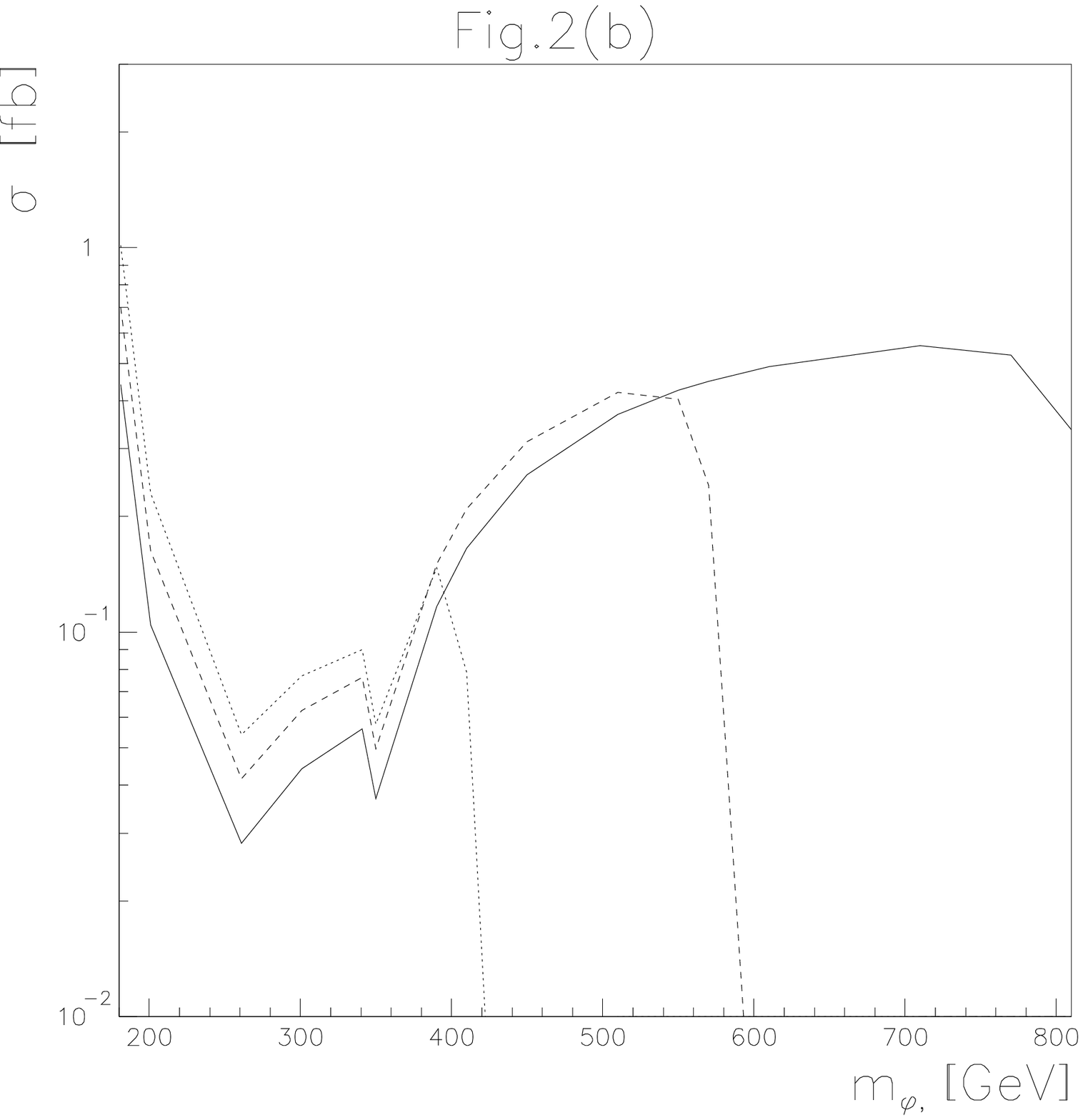}}\\
\mbox{\epsfxsize=6.truecm\epsfysize=6.truecm\epsffile{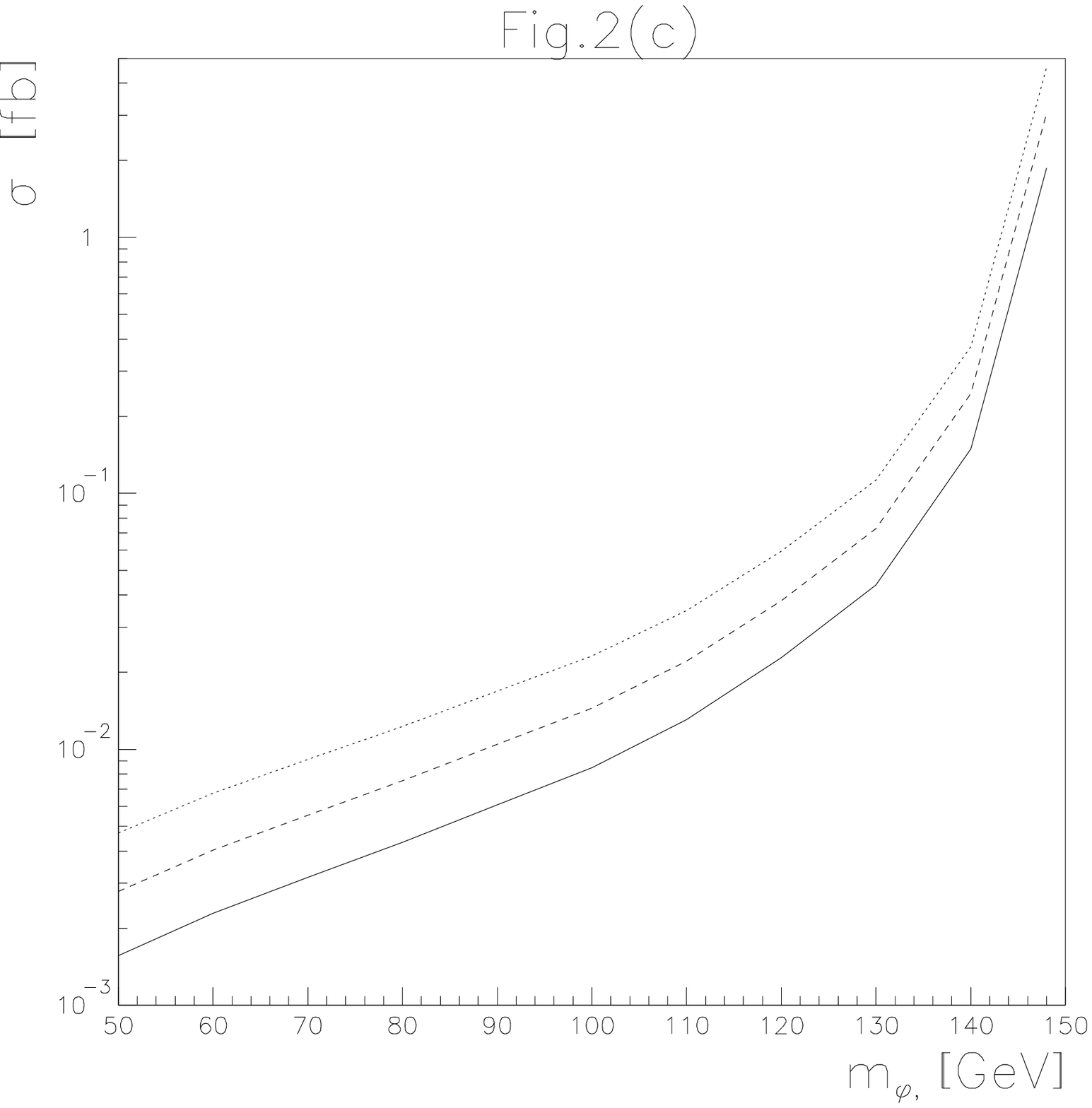}}
\mbox{\epsfxsize=6.truecm\epsfysize=6.truecm\epsffile{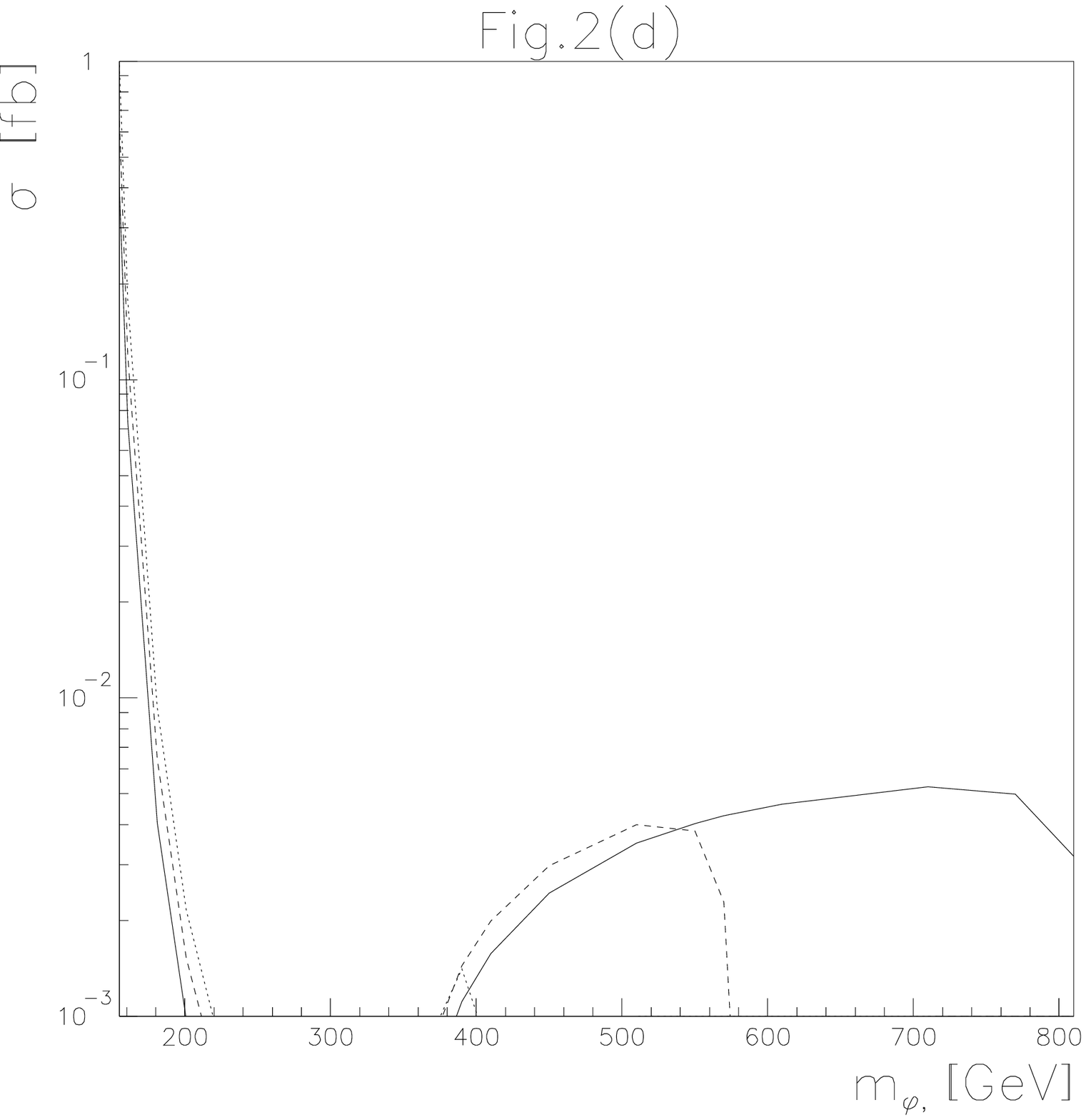}}
\end{center}
\caption{\label{fig2a} Cross section of 
$\gamma\gamma \rightarrow\phi^{'}$
        as a function of $m_{\phi^{'}}$
        with   $m_h=150$ GeV and $\xi=1/6$.
In (a) and (b) $\Lambda_{\phi}=1$ TeV, and in 
(c) and (d) $\Lambda_{\phi}=10$ TeV.
The solid line corresponds to $E_{cm}=1$ TeV, the dashed line
to 700 GeV, and the dotted line 500 GeV. }
\end{figure}
In Figs. 2 (a) and (b), we plot the
cross section of $\gamma\gamma \rightarrow
\phi^{'}$ as a function of the mass of the radion, $m_{\phi^{'}}$, at
c.m. energy $\sqrt{s}$=500 GeV, 700 GeV, and 1 TeV, respectively,
when radion and Higgs are mixed, $\xi=1/6$.
\footnote{
The gap in the Fig.2 for $m_{\phi ^{\prime}}$ between 130 GeV to
190 GeV in Fig. 2 is due to the discontinuity in $a^2$ and $c^2$
in Eq.(2.10).}
The other parameters are set to 
$m_h=150$ GeV, and $\Lambda_{\phi}=1$ TeV.
We see from Figs. 2 (a) and (b) that there is an enhancement 
when $m_{\phi^{'}}$ is near $m_h$ and a suppression
when $m_{\phi^{'}}$ is much heavier than $m_h$.
In Fig. 2 (c) and (d), we plot the cross section 
with otherwise the same parameters, but with $\Lambda_{\phi}=10$ TeV.
We find similar enhancement and suppression than in Figs. 2 (a) and
(b),  but much stronger. That is reasonable because mixing with Higgs 
will effect radion stronger with raising of $\Lambda_{\phi}/v$. 

In order to compare with
the possible suppression and enhancement of mixing in  $e^+e^-$
collision, we draw the parameter
$a_{12}$ of Eq. 2.12 as a function of $m_{\phi^{'}}$
for $\xi=1/6,m_h=150$ GeV and $\Lambda_{\phi}=1$ TeV and 10 TeV,
respectively, in Fig. 3 (a) and (b). 
Compared with \cite{s5b}, the cross section from $e^+e^-$ 
collision is in our case smaller by a factor of $10^{-4}$  for 
radions heavier than 500 GeV but larger by two orders of magnitude 
for radions with mass $m_h$.  
\begin{figure}[t]
\leavevmode
\begin{center}
\mbox{\epsfxsize=6.truecm\epsfysize=6.truecm\epsffile{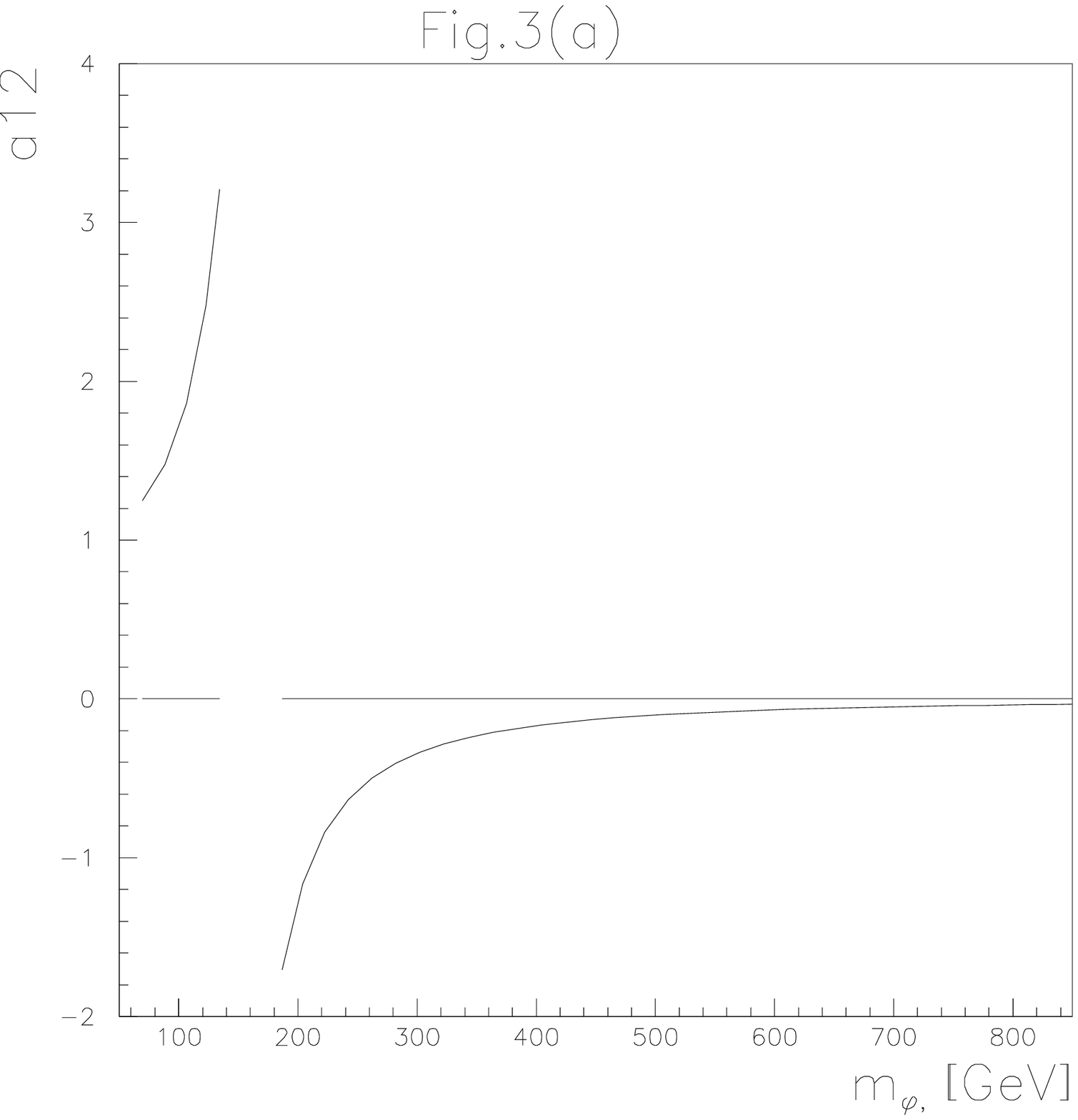}}
\mbox{\epsfxsize=6.truecm\epsfysize=6.truecm\epsffile{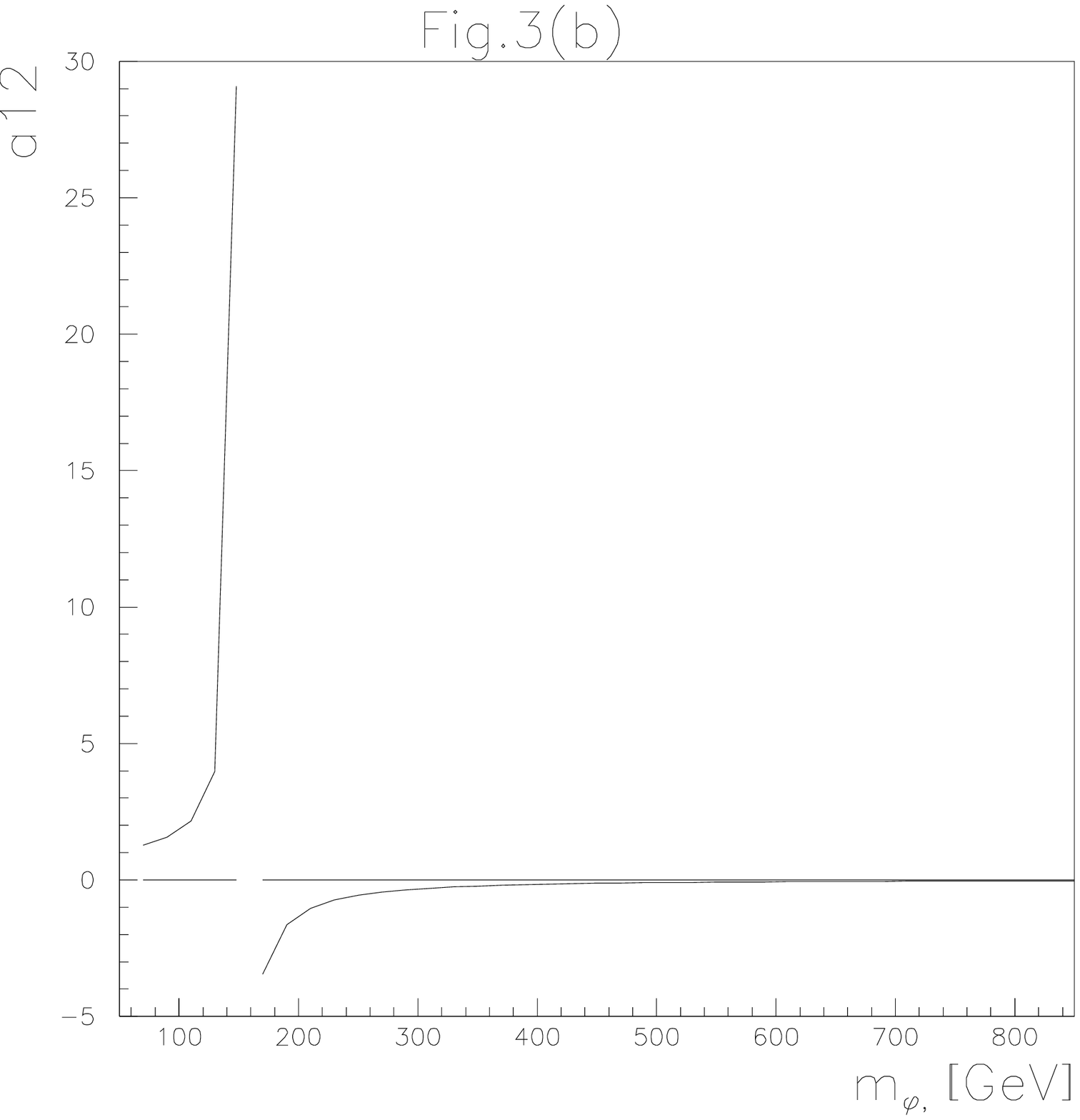}}\\
\mbox{\epsfxsize=8.truecm\epsfysize=6.truecm\epsffile{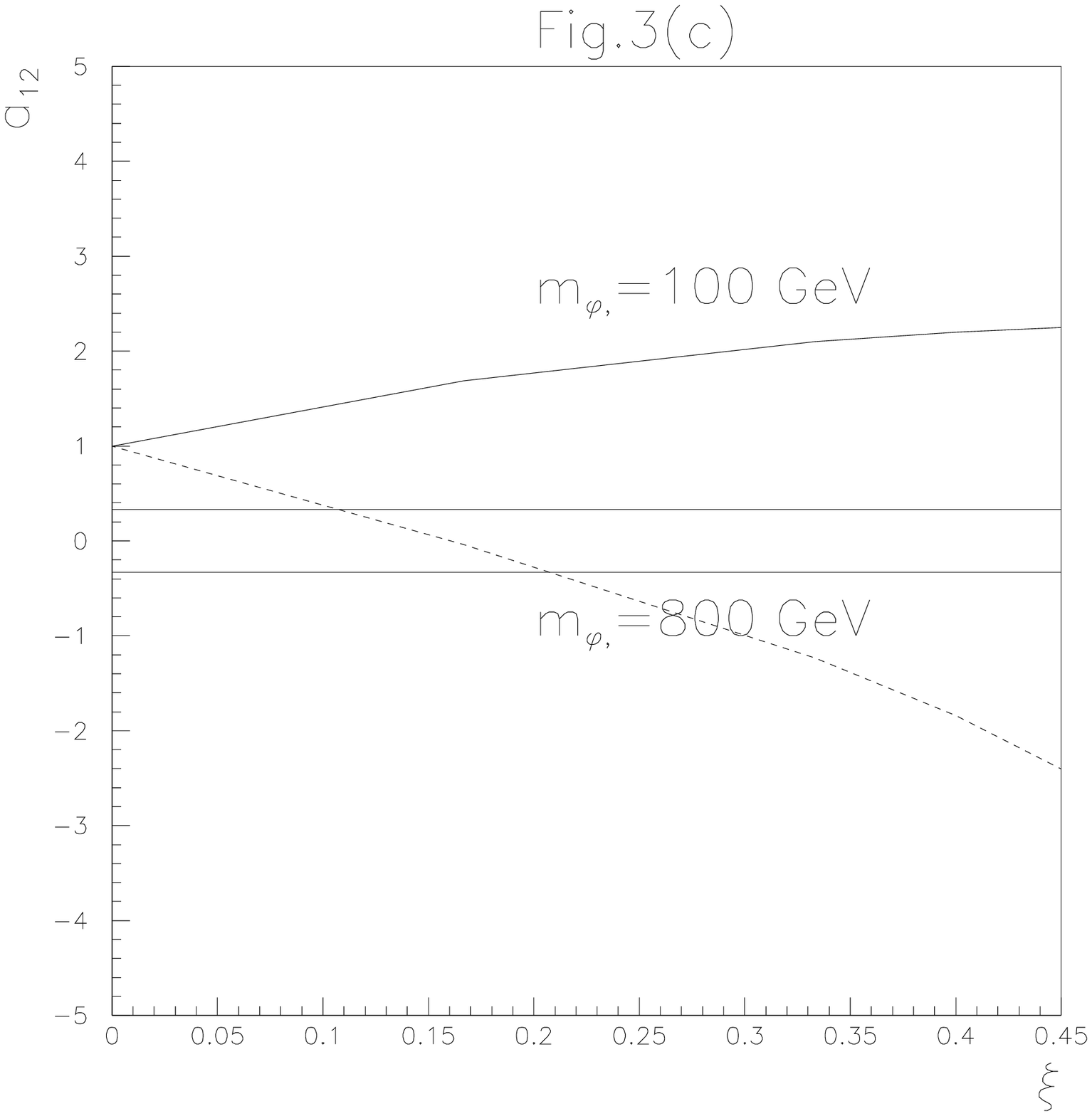}}
\end{center}
\caption{\label{fig4a}  $a_{12}$
        as a function (a) of $m_{\phi^{'}}$
        with  $\Lambda_{\phi}=1$ TeV and $\xi=1/6$,
                       b)  of $m_{\phi^{'}}$
        with  $\Lambda_{\phi}=10$ TeV and $\xi=1/6$ and 
(c) of $\xi$
        with  $\Lambda_{\phi}=1$ TeV.
 Solid line corresponds to $m_{\phi^{'}}=100$ GeV and dashed line to 
$m_{\phi^{'}}=800$ GeV, and the horizontal lines correspond to 
$\pm 0.33$.  
In all cases $m_h=150$ GeV.}
\end{figure}

Therefore, for light radions, the production cross section in $e^+e^-$
collisions is larger than in $\gamma\gamma$ collisions, while for
heavy radion at conformal limit the situation is the opposite.
Thus we can observe the radion with mass
larger than 500 GeV (up to 800 GeV) with $\gamma\gamma$ collisions
at LC with $\Lambda_{\phi}$ about 1 TeV (corresponding to
hundreds of events with luminosity about 500 fb$^{-1}$), and for a 
light radion $e^{+}e^{-}$ collision can scan it around the conformal limit. 

To consider a general  $\xi$, we 
draw $a_{12}$ as a function of $\xi$ in Fig. 3 (c) with $m_h=150$ GeV and
$m_{\phi^{'}}=100$ GeV and 800 GeV, respectively. We find that
around the conformal limit, in the region about $0.1<\xi<0.2$,
there will be suppression effects stronger than one order
from mixing for heavy radions, while away from the region, effects will
weaken and change to enhancement effects.
For a light radion the interaction with fermions and gauge bosons is
always enhanced due to mixing.

The cross section in the case of negative $\xi$ is given in
Figs. 4 (a) and (b),
with $m_h=150$
GeV and $\Lambda_{\phi}=1$ TeV. Instead of suppression at heavy
radions, there is a suppression for light radions and enhancement
for heavy radions. 
This is reasonable because $a_{12}\sim \left(1-
\frac{6\xi m_{\phi}^2}{m_{\phi}^2-m_{h}^2}\right)$ \cite{s6} when
$\Lambda_{\phi} >> v$.
The gap in $m_{\phi^{'}}$ for the region $m_{\phi'}\sim m_h$ is due 
to the same reason
than in Fig. 2.
\begin{figure}[t]
\leavevmode
\begin{center}
\mbox{\epsfxsize=6.truecm\epsfysize=6.truecm\epsffile{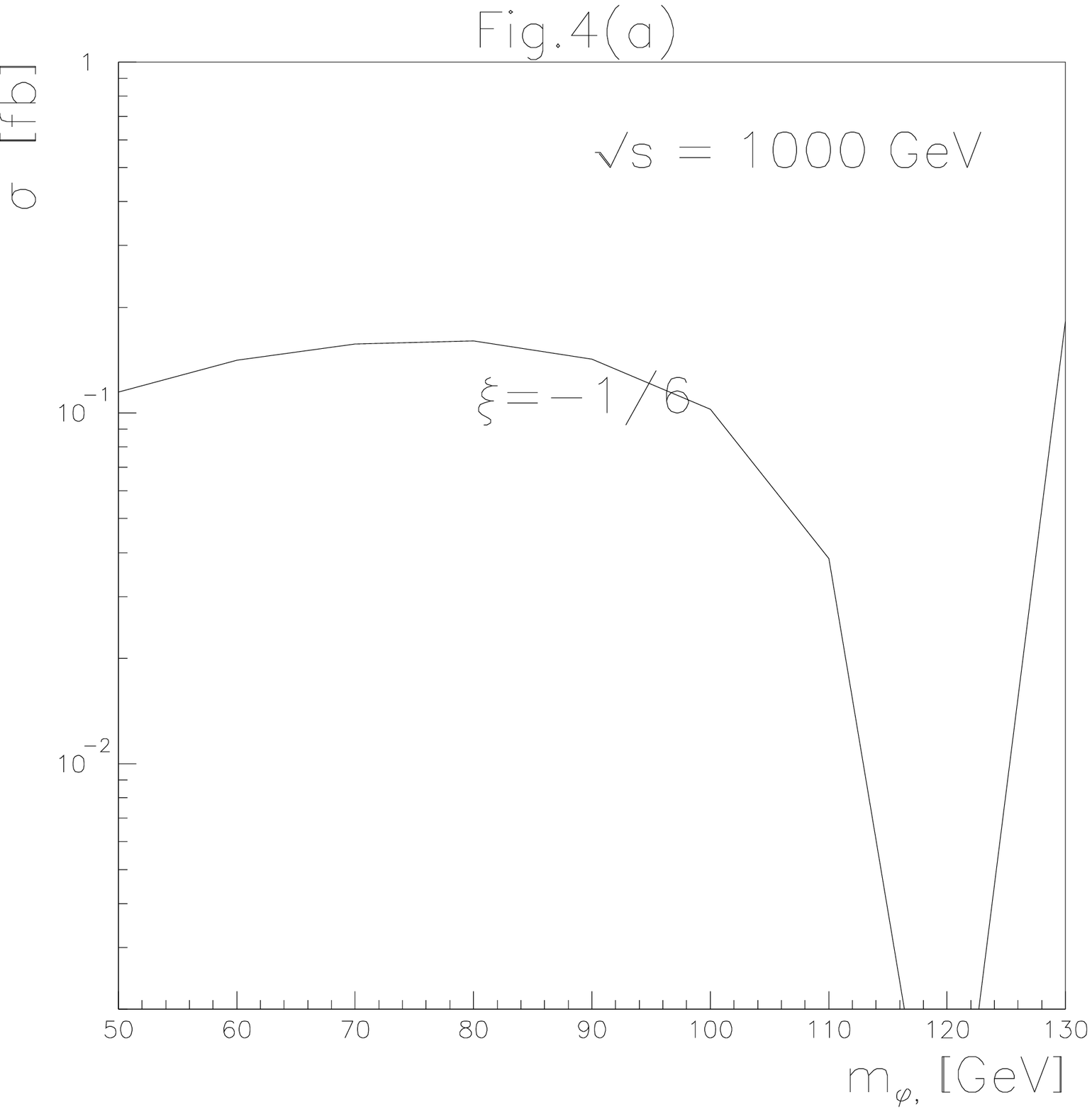}}
\mbox{\epsfxsize=6.truecm\epsfysize=6.truecm\epsffile{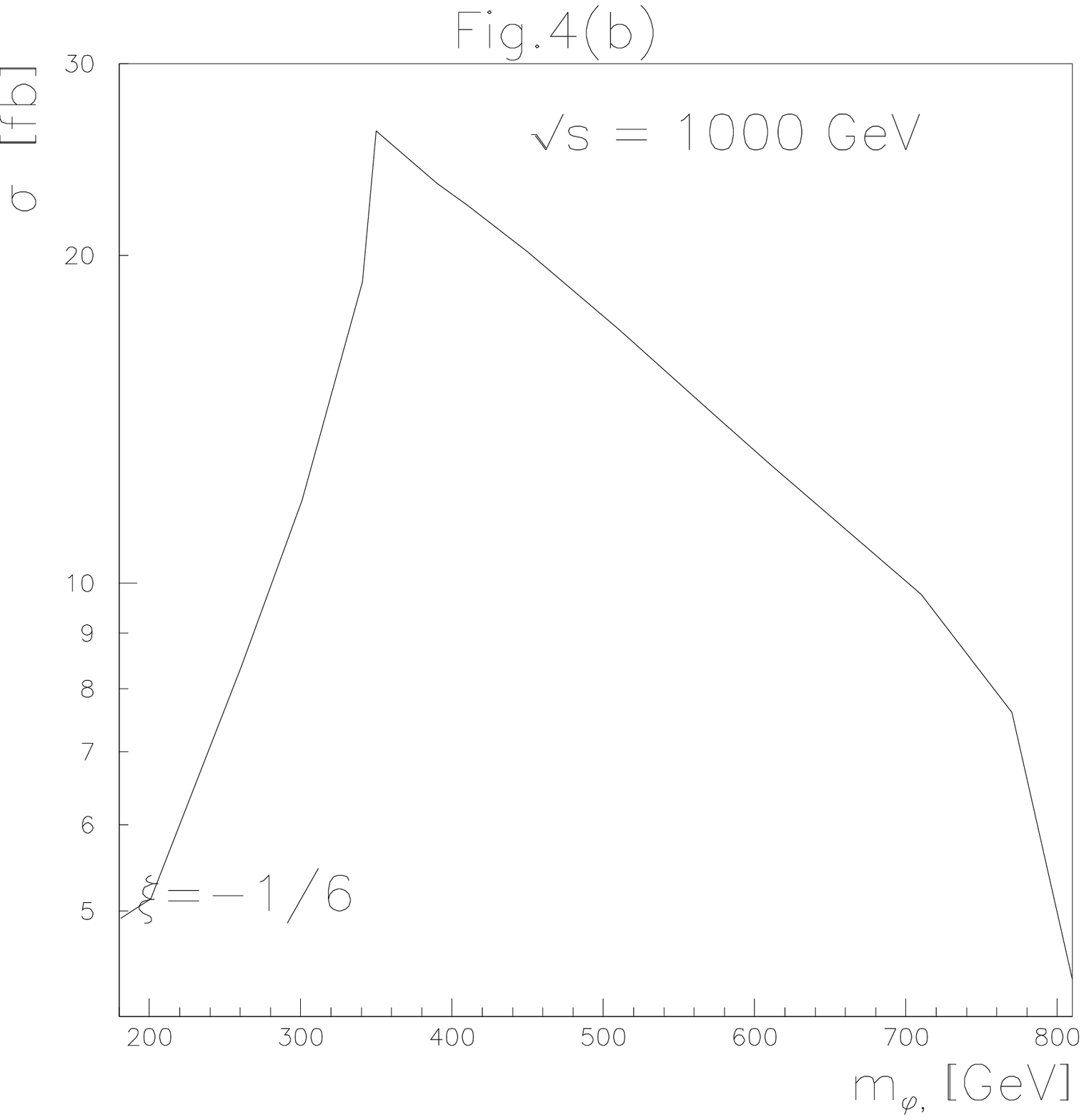}}
\end{center}
\caption{\label{fig6a} Cross section of $\gamma\gamma \rightarrow
\phi^{'}$
        as a function of $m_{\phi^{'}}$
        with  $\Lambda_{\phi}=1$ TeV, $m_h=150$ GeV, $\xi=-1/6$,
 and $E_{cm}=1$ TeV. }
\end{figure}

If the mixing of radion and Higgs is not very large and 
vacuum expectation value of radion is much larger than the VEV of Higgs,
$\Lambda_{\phi}>> v$, the mixing will effect Higgs 
considerably only when $m_{\phi} \sim m_{h}$, in which case
$a_{34}$ will be somewhat suppressed, as seen in Fig. 4 (a).
When radion is clearly heavier or lighter that Higgs,
$a_{34}$ will be near one and Higgs coupling
to fermions and massive gauge fields changes only slightly.
However, this conclusion does not hold when we consider the case
$\Lambda_{\phi}\sim
v$. In Fig. 4 (b) we draw $a_{34}$ as a function of the radion mass
with $\Lambda_{\phi}=250$ GeV, $m_h=150$ GeV, and $\xi=-1/6$. In this case
the Higgs vacuum expectation value is of the same order than the
radion, and mixing can effect Higgs strongly.
\begin{figure}[t]
\leavevmode
\begin{center}
\mbox{\epsfxsize=7.33truecm\epsfysize=5.5truecm\epsffile{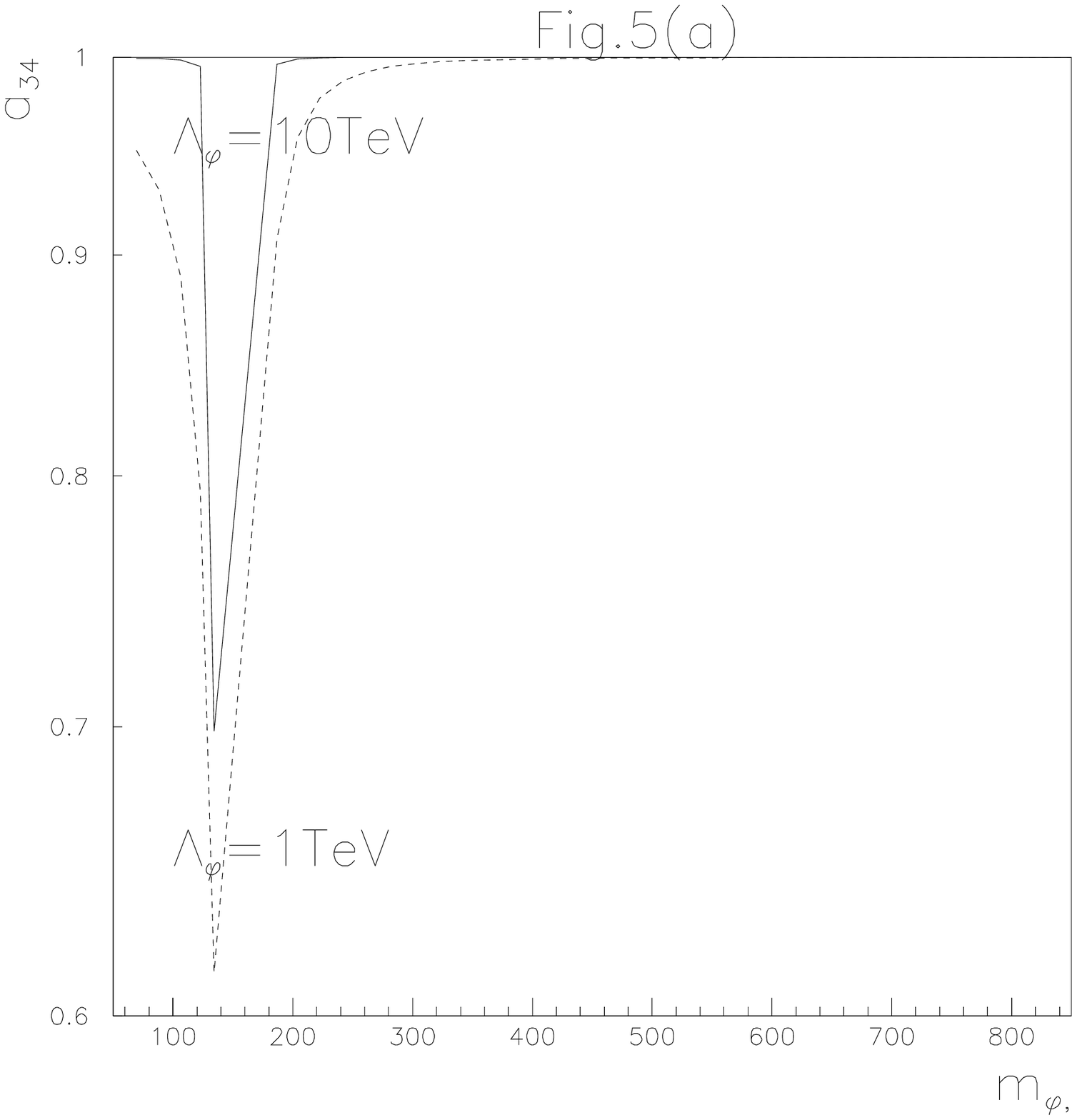}}
\mbox{\epsfxsize=7.33truecm\epsfysize=5.5truecm\epsffile{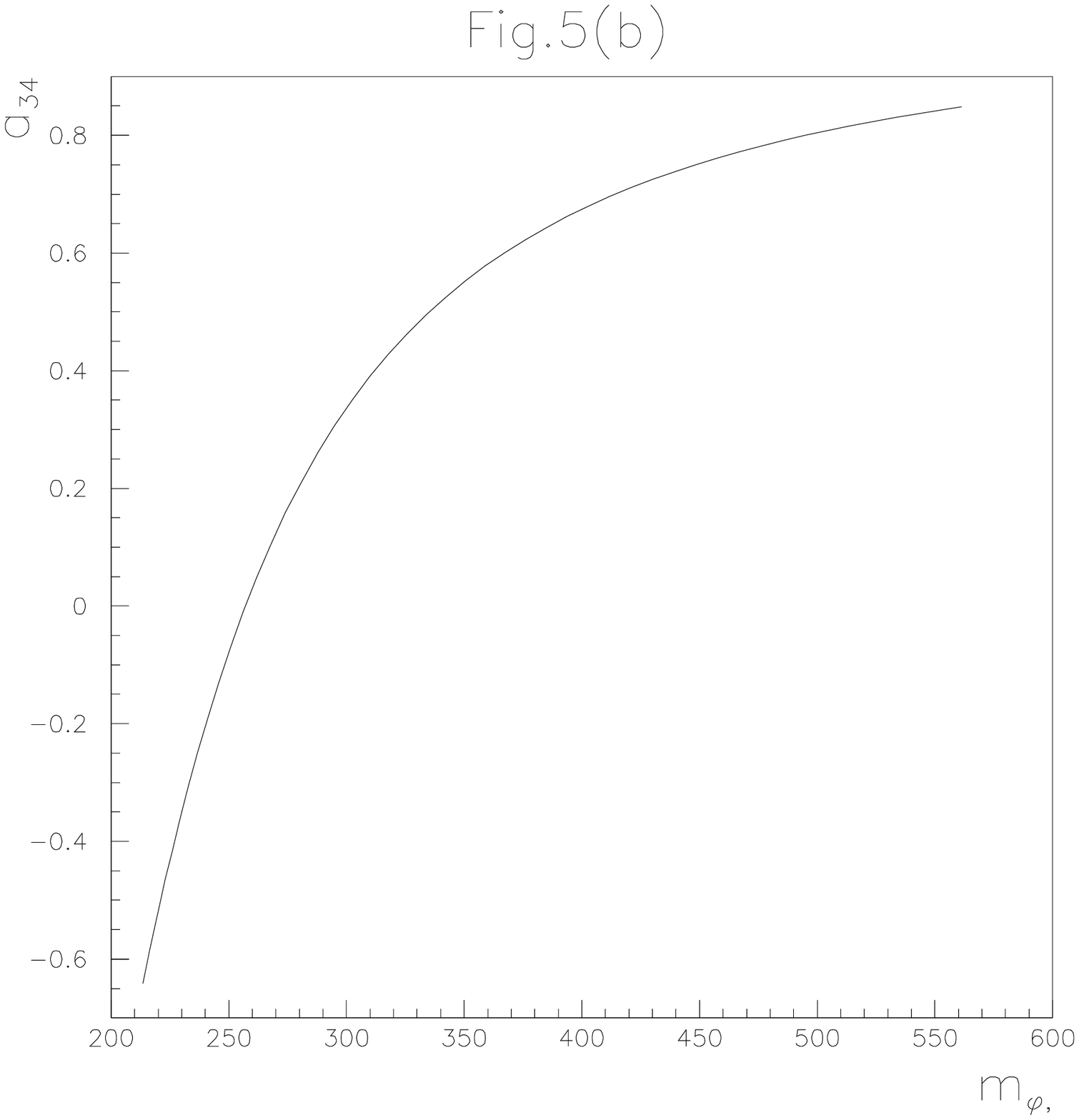}}
\end{center}
\caption{\label{fig7} $a_{34}$ as a function of $ m_{\phi^{'}} $
with $m_h=150$ GeV and $\xi=1/6$. (a) The 
solid line corresponds to $\Lambda_{\phi}=10$ TeV and the
dashed line to $\Lambda_{\phi}=1$ TeV.
(b) Here $\Lambda_{\phi}=250$ GeV, $m_h=150$ GeV and $\xi=1/6$.}
\end{figure}

\begin{flushleft} {\bf III. Radion decay} \end{flushleft} 

\noindent
In order to detect the possible radion signal at LC,
we consider the decay of radions in this section.

 In Fig.6 (a), we give the branching ratio of radion in the
no-mixing case, 
$\xi=0$, with $ m_h=150$ GeV and $\Lambda_{\phi}=1$ TeV. 
We find that 
our results
disagree with the calculations in \cite{s5b} and \cite{s6}, 
but agree with those in \cite{s5a}. 
The branching ratios for 
$\phi^{'}\rightarrow gg$ and $\phi^{'}\rightarrow b\bar{b}$ 
are both important for light radions (with mass below $2 m_{W,Z}$). 
For
heavy radions, $WW,ZZ$ will dominate and distinguishing them from Higgs
will be difficult. 
\begin{figure}[t]
\leavevmode
\begin{center}
\mbox{\epsfxsize=8.truecm\epsfysize=6.truecm\epsffile{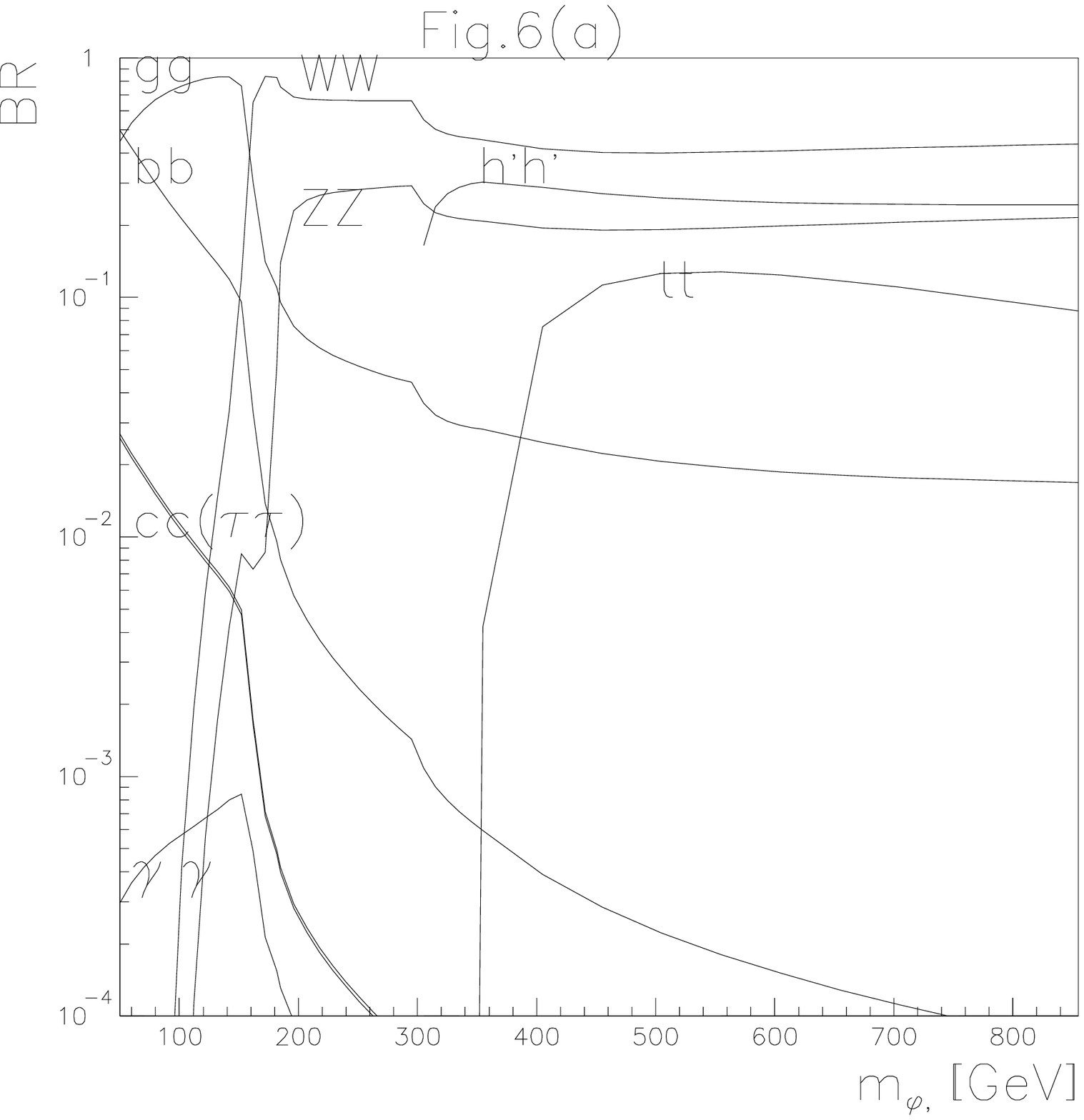}}\\
\mbox{\epsfxsize=6.truecm\epsfysize=6.truecm\epsffile{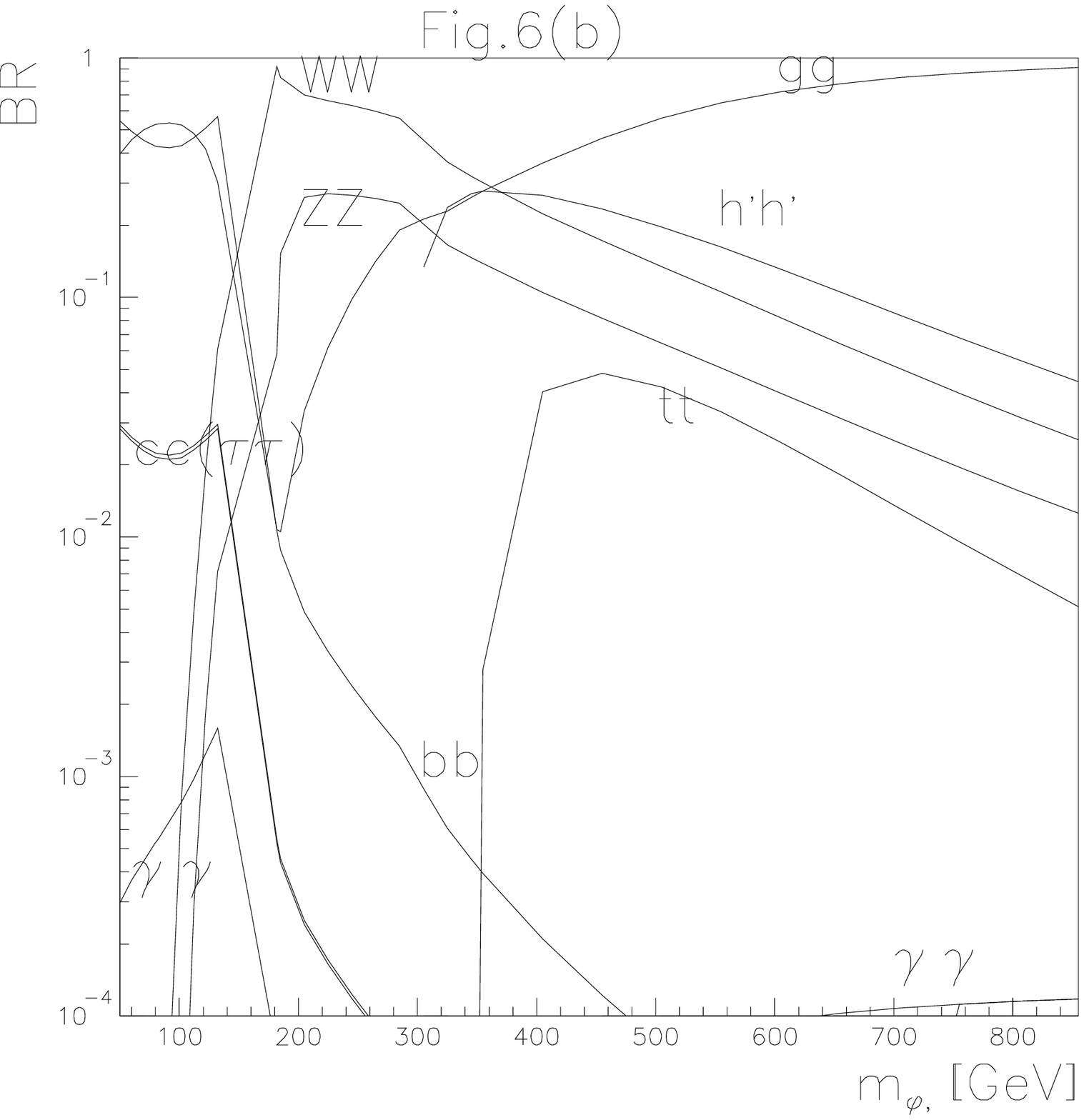}} 
\mbox{\epsfxsize=6.truecm\epsfysize=6.truecm\epsffile{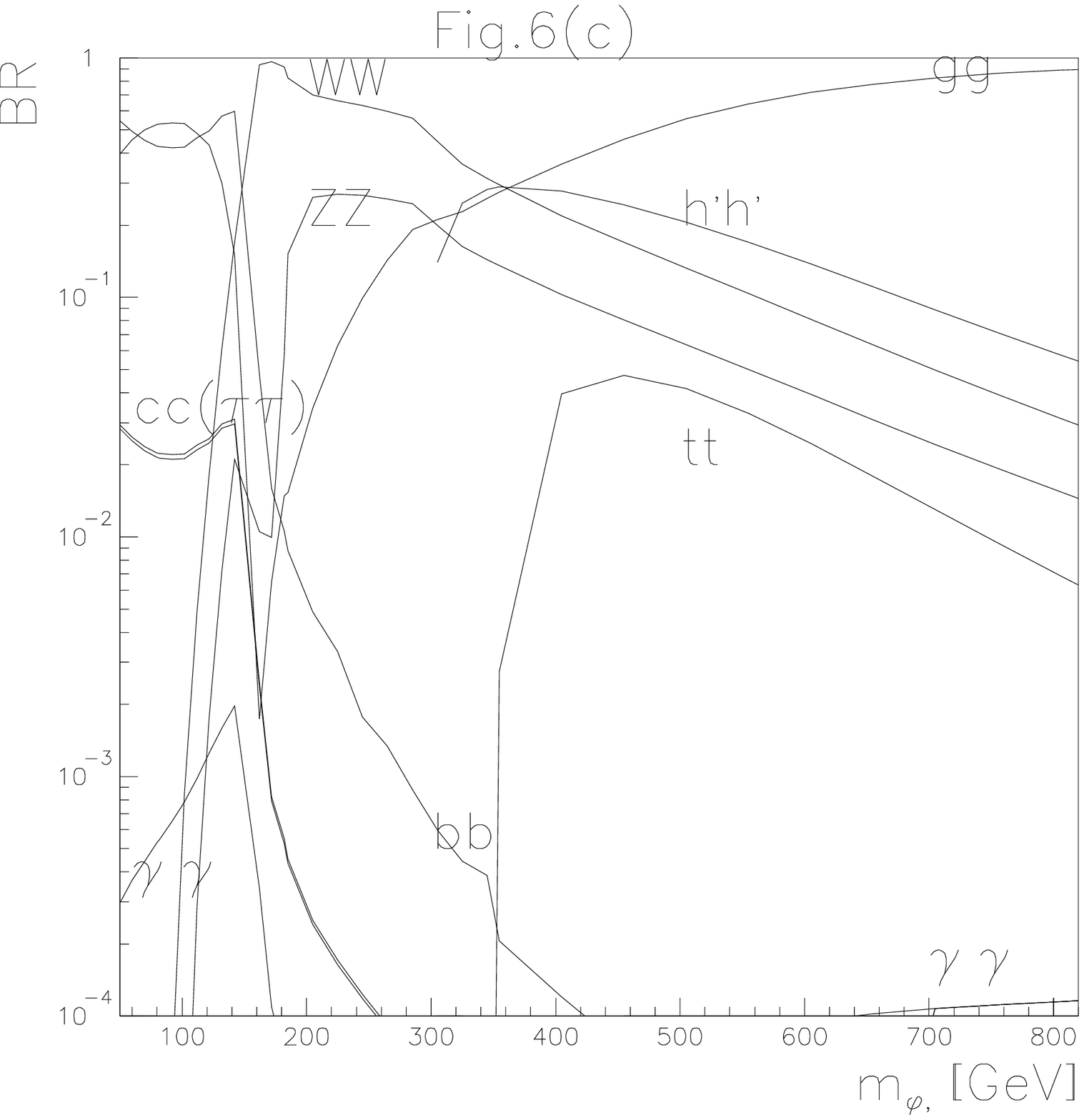}}  
\end{center}
\caption{\label{fig9} Branching ratio of radion as 
        as a function of $m_{\phi^{'}}$
        with (a) $\Lambda_{\phi}=1$ TeV and $\xi=0$.
 In
	(b) $\Lambda_{\phi}=1$ TeV and $\xi=1/6$ and
        in (c) $\Lambda_{\phi}=10$ TeV and $\xi=1/6$.
In all cases $m_h=150$ GeV.}
\end{figure}

In Fig. 6 (b) and (c), we plot the branching ratio of radion with
$\xi=1/6,\, m_h=150$ GeV and $\Lambda_{\phi}=1$ TeV, 10 TeV
respectively. In
this case, the decay of a heavy radion will be dominated by
$\phi^{'}\rightarrow gg$ and will be easily distinguished from Higgs, 
and a light radion will decay into $\phi^{'}\rightarrow gg$ and
$\phi^{'}\rightarrow b\bar{b}$, 
with approximately the same branching ratios.
Another interesting thing is that BR of $\phi^{'} \rightarrow 
\gamma\gamma$ will increase with increasing the mass of a heavy radion.

\begin{flushleft} {\bf IV. Conclusion} \end{flushleft}

\noindent
We have studied the process $\gamma\gamma \rightarrow \phi^{'}$ in
the RS model with curvature-Higgs mixing. 
The calculations show that radion production from
$\gamma\gamma$ collision will be an important way to produce 
radion at LC and even a dominating way when the mixing parameter
$\xi$ is around the conformal limit. 

Our calculations show that it should be possible to
detect a radion for $E_{cm}=1$ TeV with mass below $800$ GeV if 
$\Lambda_{\phi}$ is around 1 TeV.  Even if $\Lambda_{\phi}$ is about 10 
TeV, the light radion with mass around $m_h$ may be observable
because of enhancements from mixing.

The decay modes of radion in the mixing case will be quite different from
no-mixing case when radion is heavy or its mass is around $m_h$.
Not only a light radion, but also a heavy radion will decay mainly to
gluon-gluon, if mixing is around conformal limit.
It can be clearly distinguished from the Higgs decay.

\newpage

\begin{flushleft} {\bf Acknowledgement} \end{flushleft}

Z.-H. Yu thanks the World Laboratory, Lausanne, for the scholarship.
 
\vspace*{2cm}
 
\begin{center} {\bf Appendix} \end{center}
\par
A. Form factors
\par
The form factors $F_{1/2} (\tau_t)$ and $F_{1} (\tau_W)$ 
can be defined as \cite{s6}\cite{s11}
$$
\begin{array}{lll}
F_{1/2} (\tau) =-2 \tau [1+(1-\tau) f(\tau)]
\end{array}
\eqno{(A.2)}
$$
and
$$
\begin{array}{lll}
F_{1} (\tau) =2 + 3 \tau + 3 \tau (2-\tau) f(\tau)
\end{array}
\eqno{(A.2)}
$$
where $\tau_t=4m_t^2/q^2$, $\tau_W=4m_W^2/q^2$ and
$$
\begin{array}{lll}
f(\tau) &=&[sin^{-1} (1/\sqrt{\tau})]^2,  \ \ \ \tau \ge 1, \\
&& -1/4 [Log (\eta_{+}/\eta_{-}) - i \pi]^2, \ \tau <1,
\end{array}
\eqno{(A.3)}
$$
with $\eta_{\pm}=1\pm \sqrt{1-\tau}$.
\par
B. $\gamma\gamma$ collision
\par
  In order to get the observable results in the measurements of
radion production via $\gamma \gamma$ fusion
in $e^{+}e^{-}$ collider, we need to fold the cross section of
$\gamma\gamma
\rightarrow \phi^{'}$ with the photon luminosity,
$$
\sigma(s) = \int_{m_{\phi^{'}}/\sqrt{s}}^{x_{max}} dz
\frac{dL_{\gamma\gamma}}{dz}
              \hat{\sigma}(\hat{s}),
\eqno {(B.1)}
$$
where $\hat{s}=z^2 s$, $\sqrt{s}$ and $\sqrt{\hat{s}}$ are the
$e^{+}e^{-}$
and $\gamma\gamma$ c.m. energies respectively, and
$\frac{dL_{\gamma\gamma}}{dz}$
is the photon luminosity, which is defined as \cite{s7} 
$$
\frac{dL_{\gamma\gamma}}{dz} = 2z \int_{z^{2}/x_{max}}^{x_{max}}
\frac{dx}{x}
                                F_{\gamma /e}(x)F_{\gamma /e}(z^{2}/x).
\eqno {(B.2)}
$$
The energy spectrum of the back-scattered photon is given by \cite{s7}.
$$
F_{\gamma /e}(x) = \frac{1}{D(\xi)}[1-x+\frac{1}{1-x}-\frac{4x}{\xi
(1-x)}+
                   \frac{4x^{2}}{\xi^{2} (1-x)^{2}}].
\eqno {(B.3)}
$$
taking the parameters of Ref. \cite{s12}, we have
$\xi=4.8$, $x_{max}=0.83$ and $D(\xi)=1.8$.

\end{document}